# (Mis)Information Operations: An Integrated Perspective

M Cinelli[1], M Conti[2], L Finos[2], F Grisolia[3], P Kralj Novak[4], A Peruzzi[5], M Tesconi[7],  F Zollo[5][6], W Quattrociocchi[5]

[1]ISC-CNR,
Via dei Taurini 19, 00185
Rome, Italy

E-mail: matteo.cinelli@roma1.infn.it

[2]University of Padua
Padua, Italy

E-mail: conti@math.unipd.it; livio.finos@unipd.it

[3]University of Pisa
Pisa, Italy

E-mail: francesco.grisolia@di.unipi.it

[4]Jožef Stefan Institute
Ljubljana, Slovenia

E-mail: petra.kralj.novak@ijs.si

[5]Ca' Foscari University of Venice
Venice, Italy

E-mail: w.quattrociocchi@unive.it; fabiana.zollo@unive.it;
antonio.peruzzi@stud.unive.it

[6]Center for the Humanities and Social Change
Venice, Italy

E-mail: fabiana.zollo@unive.it

[7]IIT-CNR
Pisa, Italy

E-mail: maurizio.tesconi@iit.cnr.it

**Abstract:** *The massive diffusion of social media fosters disintermediation and changes the way users are informed, the way they process reality, and the way they engage in public debate. The cognitive layer of users and the related social dynamics define the nature and the dimension of informational threats. Users show the tendency to interact with information adhering to their preferred narrative and to ignore dissenting information. Confirmation bias seems to account for users' decisions about consuming and spreading content; and, at the same time, aggregation of favored information within those communities reinforces group polarization. In this work, the authors address the problem of (mis)information operations with a holistic and integrated approach. Cognitive weakness induced by this new information environment are considered. Moreover, (mis)information operations, with particular reference to the Italian context, are considered; and the fact that the phenomenon is more complex than expected is highlighted. The paper concludes by providing an integrated research roadmap accounting for the possible future technological developments.*

## Introduction

Information operations make use of information-related capabilities to influence, disrupt, corrupt, or usurp the decision-making of a target audience (Joint Chiefs of Staff 2014). These operations can include a combination of methods, such as false news, disinformation, or networks of fake accounts (false amplifiers), with the aim of manipulating public opinion. In this paper, the authors refer to (mis)information operations as the specific and organized use of misleading information to influence a target audience. While, on the strategic level, the aim is to influence public opinion and social perceptions on specific issues in order to alter the process of decision-making, on the tactical level, (mis)information operations aim to trigger specific social responses (such as encouraging consumers to share such content, vote certain ways, or participate in online debate) through the use of provocative content (Wanless & Berk 2019). Since the effect of (mis)information operations relies on the exploitation of social responses to specific issues, their strategies, tactics, and operational implementation may change together with social and technological developments.

Currently, the massive diffusion of sociotechnical systems and micro blogging platforms on the World Wide Web (WWW) created a new scenario for information consumption. This information ecosystem grounded on disintermediation—that is, the direct access to content through social media platforms—can be exploited by (mis)information operations. The new interaction patterns allow for a direct path from producers to consumers of content, and change the way users are informed, engage in debate, and develop opinions (Del Vicario *et al.* 2016b). This shift, in turn, affects political communication (Stieglitz, Brockmann & Dang-Xuan 2012) and the evolution of the public debate (Bond *et al.* 2012).

A multitude of mechanisms influence the way in which people share and absorb information. For example, the process of acceptance of a claim (whether documented or not) may be affected by normative social influence or by coherence with an individual system of beliefs. The problems arising from this renewed vulnerability to (mis)information operations do not seem to be well defined; indeed, several scientific works have produced conflicting definitions, results, and insights both on a quantitative and a qualitative basis (Ruths 2019). Some think that the consequences of this disintermediated environment could be reduced to a fallacious antagonism: truth against false. However, social media dynamics are complex, and different factors may come into play when dealing with online human cognitions and preferences. Supporters of this dichotomous standpoint often forget that social media are mainly intended for ludic rather than for informative purposes (De Waal & Schoenbach 2010) and that news consumption on social media is often incidental rather than deliberate (Boczkowski, Mitchelstein & Matassi 2018). For example, the most liked image on Instagram is an egg (54 million likes).

In recent years, the concern about (mis)information has grown in importance. In 2017, the World Economic Forum raised a warning on the potential distortion effect of social media on the perception of reality (World Economic Forum WEF 2017). The challenge is pressing not only at the scientific level (Del Vicario *et al.* 2016b; Quattrociocchi 2017; Sunstein 2018; Schmidt, AL *et al.* 2017, Schmidt, AL *et al.* 2018b; Grinberg *et al.* 2019) but also at the political level to an extent that governments of Western democracies are pursuing initiatives to limit such a tendency, and social media companies are operating plans to counteract potential manipulation (Schmidt, C 2019). The stakes are high, particularly in democracies where trust in information is paramount to legitimacy of the system (Bimber 2003).

However, the emerging solutions are not likely to be effective, especially those adopting the above-mentioned dualistic approach. To this respect, an approach widely adopted by major news outlets focuses on contrasting misleading content by means of fact-checking news pieces. To understand why this approach is fallacious requires an understanding of the underlying dynamics of social media. Along this path, quantitative studies showed the inner tendency of users to interact with information adhering to their preferred narrative (Del Vicario *et al.* 2016b) and to ignore dissenting information, for example in the case of debunking pieces (Zollo *et al.* 2017). In this respect, confirmation bias was shown to play a role in influencing users' decisions about consuming and spreading content. This mechanism fosters the creation of echo chambers, self-segregated groups of users clustering around and reinforcing shared narratives, while rejecting any possible alternative point of view (Quattrociocchi 2017; Sunstein 2018).

The role played by automation and the deliberate manipulation of online algorithms further exacerbates the situation. Heavily automated Twitter posts have been found to influence online news feeds and search returns (Mustafaraj & Metaxas 2010). Also, online search results have been found to affect political opinion (Epstein & Robertson 2015). Thus, it is not surprising that the scope of (mis)information operations are pursued and amplified by exploiting opportunities and vulnerabilities of web platforms. For instance, Internet bots (Vosoughi, Roy & Aral 2018), which can also exploit social network platforms (Compatno *et al.* 2015), have been found to play a role in increasing the speed of information spreading that is still hard to quantify in a precise way. In spite of these aspects, there is enough evidence to suggest that bots foster misinformation spreading in a manner leading the attitudes of some people to be more prone to engage with fake news than others (Guess, Nagler & Tucker 2019; Grinberg et al. 2019) and in influencing election/referendum outcomes and society as a whole (Grinberg *et al.* 2019, Vosoughi, Roy & Aral 2018, Del Vicario *et al.* 2017).

By reducing the phenomenon just to the dichotomous distinction between 'trustworthy' and 'unreliable' content, the overall picture is not being captured; thus, researchers are answering badly flawed questions. This false dichotomy is a problem since it oversimplifies the issue about the consequences of (mis)information operations in this new disintermediated information environment, drawing a line in the sand that is often difficult to discern, particularly at scale, but nearly impossible to distinguish when it comes to something like political opinion; such an approach is reductionist and fails to take into account the many cognitive issues that (mis)information operations deliberately provoke. To understand and hopefully address the consequences of (mis)information operations in this changing information environment, the tendency to seek information adhering to one's worldview and to ignore dissenting information must be remembered. No matter whether offline or online, people tend to live and express themselves inside so-called echo chambers. (Sunstein 2002; Quattrociocchi 2017). Unfortunately, confirmation bias is inherent in human cognition and is not likely to be eliminated (Nickerson 1998). The good news, and the challenge, is that addressing misinformation might become possible through a multi-disciplinary collective approach.

The purpose of this work is to improve the understanding of the problem of (mis)information diffusion in the current disintermediated information environment by embracing the complexity of human cognitions and interactions. The first section is devoted to the underlying social phenomena through the lens of cognitive and clinical psychology. In particular, the focus is on the role played by confirmation bias—with the consequent emergence of polarized groups of users—and on the relevance of specific personality traits in shaping the way social media users absorb and interact with information. In the second section, the focus is on the ecosystem of the new information environment and on the proliferation, within this environment, of automated ways of exploiting human vulnerabilities—along with algorithmic flaws—for (mis)information-operation purposes. Not only a brief overview of the social media ecosystem will be given, but also the main typologies of social bots will be described. The third section will deal with three Italian examples of misinformation in the political sphere, explored through a cognitive lens. The fourth section will be devoted to the discussion of a possible road map for further research on the topic. The main points of this work will be summarized in the last section.

## Users Behavior through the Lens of Cognitive and Clinical Psychology

As already mentioned, it is necessary to start from the dynamics of social media in order to properly address the rising problem of (mis)information diffusion. These dynamics are not caused by social media but are the effects of human cognition operating in the media environment. For this reason, the field of cognitive psychology seems to offer the right set of conceptual tools to address the issue. Cognitive psychology is the branch of psychology dealing with the way people acquire, process, and memorize information (VandenBos 2007). In this case, understanding the way social media users absorb and process information online seems to be crucial.

On the Internet, a huge amount of information competes for viewers' attention, which is instead limited; this allows their cognitive biases—shortcuts or heuristics that are used to simplify reality and (re)act rapidly (Haselton, Nettle & Murray 2015)—to take the lead in processing information. Humans use such biases to interpret reality. Unfortunately, while these cognitive mechanisms are often fundamental to survival, they might also act as mental traps and mislead viewers. A crucial role in information consumption and diffusion is played by confirmation bias, which is the human tendency to look for information that is already consistent with one's system of beliefs. Indeed, online users tend to fragment into echo chambers (Del Vicario *et al.* 2016b)—polarized communi-ties whose users share a common narrative. Immersed in echo chambers, users select information consistent with their worldviews, even when false (Bessi *et al.* 2016), while ignoring information that dissents from their beliefs. Users from different and contrasting communities rarely interact; and, when that happens, the debate degenerates, especially for longer discussions (Zollo *et al.* 2015). Response to debunking attempts is not that dissimilar and results in the well-known 'back-fire effect', thus reinforcing users' original positions (Zollo *et al.* 2017).

It is possible to quantify the turnover of Facebook news sources by measuring the heterogeneity of users' activity. It may be observed that, for increasing levels of activity (number of likes) and 'lifetime' (the temporal distance between the first and last interaction of a user to a post on the platform), users interact with increasingly fewer new sources (Schmidt, AL *et al.* 2017). News consumption on Facebook is therefore dominated by selective exposure (Cinelli 2019), showing a natural tendency of users to confine their activity to a limited set of pages, while focusing their attention on certain topics (and claims). This self-selection contributes to the formation of a highly-polarized community structure. Such dynamics appear to be independent of the topic, and also apply to online political debates, such as the ones around Brexit or the Italian Constitutional Referendum's debate (Del Vicario *et al.* 2017). Users' segregation in echo chambers may play a pivotal role in the spread of information on social media.

To contrast misinformation, and encourage effective communication, smoothing polarization is thus essential. To this end, users' behavior and their interactions with information may be used for a timely identification of potential misinformation targets (Del Vicario *et al.* 2019). This could allow for the design of tailored counter-narratives and appropriate communication strategies. The EU H2020 project QUEST (https://questproject.eu) aims at analyzing, designing, testing, and evaluating different strategies to improve science communication on social media, with a special focus on delicate and polarizing topics such as climate change or vaccines (Schmidt, AL *et al.* 2018b). Along the same line, a recent investigation (Pomerantsev *et al.* 2019) analyzed the engagement of *Corriere* (a newspaper) readers with content touching on the controversial topic of migration in Italy. The purpose was to understand which types of journalism intensify or reduce polarization and identify the best way to communicate facts and foster constructive engagement. Findings indicate that impartial, accurate reporting draws the most institutional trust, while human interest stories can encourage strong negative comments from readers. Consistent with previous results, infographics, fact-checking, and a data-driven approach may elicit strong pushback and criticism from audiences and thus boost polarization.

Another aspect that deserves attention seems to be the predominance of specific psychological traits in the social media environment. Users exhibiting certain traits may be, in fact, more vulnerable than others in promoting content which is the object of (mis)information operations. Such an approach exploits tools borrowed from clinical and personality psychology, which is the branch of psychology assessing the overall behavioral and health issues of the individuals (VandenBos 2007). For example, individuals with high indicators of narcissism and lower levels of self-esteem seem to be associated with greater online activity (Mehdizadeh 2010). Moreover, the lack of emotional regulation and coping results seems to be strictly associated with problematic Internet use (Spada & Marino 2017; Marino *et al.* 2017; Marino *et al.* 2018; Marino *et al.* 2019b). Even the style of attachment to father and mother and metacognitions predict the problematic use of both Internet and social media (Marino *et al.* 2019a).

Evidence suggests a relationship between narcissistic personality disorder traits and passive data from social network sites, mainly the number of friends. A possible hypothesis for this relationship is the presence of the attention-seeking trait, which is also a core component of histrionic personality disorder. Narcissism together with aggression and lack of self-control also correlate with online game addiction (Kim *et al.* 2008; Eksi 2012). Other traits associated with a problematic usage of social media are shyness, loneliness, social anxiety, and lack of self-esteem (Chak & Leung 2004; Caplan 2006; Mehdizadeh 2010). However, these negative traits are not the only ones in play in this environment. The degree of influence in some communities seems to be more related to openness, consciousness, and emotional competence (Zanotto 2017). Psychologists and sociologists depict a very complex scenario that, while providing some answers, also leaves many questions open and stimulates new questions. A future challenge might be the prediction of the problematic traits linked to a higher diffusion of misinformative content, but more generally, the profiling of users' personalities on the basis of their (online) social activities.

## '(Mis)information Automation' in the Changing Information Environment

Another phenomenon within the changing information environment that should be addressed is the presence of automated systems exploiting both the technical flaws emerging from such an environment and human cognitive biases. This changing information environment, with its disintermediation, presents exciting new opportunities for people to communicate, but has also proven to be a noisy space, where navigating between good and bad information is challenging for the average user. This environment is quite exposed to (mis)information operations and, in particular, the presence of automated bots further exacerbating the problem of misinformation diffusion by leveraging the previously-mentioned human vulnerabilities. Before fully getting into this issue of '(mis)information automation', it seems necessary to take a closer look at the components in the ecosystem of this disintermediated environment. One such component is represented by social bots.

Several interesting phenomena can be observed on social media, such as the proliferation of political pages and alternative information sources with the aim of using the Internet to organize and convey public discontent (Mocanu *et al.* 2015). Furthermore, very distinct groups, namely trolls, have emerged and built Facebook pages as a parodistic imitation of both alternative information sources and online political activists. Their activities range from controversial and satirical posts mimicking alternative news sources to outright fake claims. These memes may go viral and are used as evidence in online debates by political activists (Del Vicario *et al.* 2016a). High levels of distrust in official institutions help the diffusion of alternative, misleading explanations, both the unsubstantiated or inaccurate, including conspiracy theories which tend to explain a significant social or political aspect as a secret plot by powerful individuals or organizations. This kind of activity is proliferating within echo chambers. To make things even more complex, recent studies have shown that the social media environment is populated by a multitude of accounts purposely created to spread unsolicited spam, advertise products of doubtful legality, sponsor public figures, or influence the public opinion (Jiang, Cui & Faloutsos 2016).

It is in this context that automation comes into play in various forms. For example, a social bot is a computer program that controls a social media account, mimicking a legitimate user, while a cyborg (Chu *et al.* 2012) blends characteristics of both manual and automated behavior. Several coordinated accounts cooperating towards a common goal constitute a social botnet. As already mentioned, these kinds of automated systems may increase the spread of misinformation by leveraging the tendencies in some people to be more vulnerable to deceptive information than others (Guess, Nagler & Tucker 2019; Grinberg *et al.* 2019). The threat is that the exploitation of these tools may lead to influencing election/referendum outcomes and society as a whole (Grinberg *et al.* 2019, Vosoughi, Roy & Aral, 2018, Del Vicario *et al.* 2017).

## (Mis)information Operations in Italy

Myriad actors are using (mis)information operations to set agendas and shape public opinion. In an environment where mediators such as mainstream media are pressed to find a new business model that allows direct access to an unprecedented plethora of sources (Cinelli *et al.* 2019), the public gets lost in the large amount of misleading information they are exposed to, and this corrupts trust in favor of partisan debates. (Mis)information operations are seldom clear cut and often the activities of actors blur in the digital space, making it difficult to discern where one begins and the other ends. The Italian political debate is rife with (mis)information operations, as explored below.

For example, the Five Star Movement (the Movement), the Italian 'non-party', founded by a famous comedian, was supported by a coordinated network of social media pages and websites falling outside the official channels of the Movement. According to Nardelli and Silverman (2016), these pages and platforms relentlessly published misinformation ranging from conspiracy theories to partisan campaigns despite their claim of being "independent sources". Their degree of independence is questionable, with some of these online holdings sharing IP addresses, Google Analytics and AdSense IDs with the Movement (Nardelli and Silverman 2016). The (mis)information operations used to support the Movement included activists sharing fake poll results indicating that the party had more support than it had at the time (Gavazzi 2018), the use of divisive issues such as immigration and vaccines (Mackay 2019) to provoke Italian audiences, and the use of fake social media accounts to amplify such content (Associated Press 2019).

Another case regarding one Facebook account that was closed highlights the blurred lines between actors attempting to shape the information environment. It was connected to media entrepreneur Giancarlo Colono, who own networks of websites (175 domains) and corresponding Facebook pages, pushing viral clickbait to "hyperpartisan pieces about immigration that echo nationalist and Islamophobic rhetoric" (Nardelli and Silverman 2017). Some of the pages owned by Colono, such as DirettaNews, have more than 3 million followers on Facebook. Colono and his family have also supported and reshared content online from the Catholic organization, *La Luce di Maria*, who propagandize the faith, and whose Facebook page shares alternative health posts and "tips to fight malevolent spells and the devil" (Nardelli and Silverman 2017). DirettaNews has been found to repost articles from *La Luce di Maria*. The *La Luce di Mari* web domain had been registered to one Roberto Granieri, whose Facebook account was an administrator on the Italians First closed Facebook group, also connected to another Colono web holding, iNews24, and "was a member of numerous public and private far-right, nationalist, anti-migrant, and anti-Islam Facebook groups, as well as pro-Putin, Five Star Movement, and nationalist Lega party leader Matteo Salvini groups" (Nardelli and Silverman 2017). While Colono claimed the Granieri account was fake, its activities and connections illustrate how a variety of political operators intersect online attempting to shape the information environment. Distinguishing one from the other can be very challenging (Nardelli & Silverman 2017).

The Italian political party *Lega Nord*, now *Lega per Salvini Premier* (*Lega*), was also found to be using botnets to shape the information environment. Activists supporting *Lega* used messaging boards such as 4chan and 8chan to radicalise and recruit youth (Ebner & Davey 2018). The social media managers of *Lega* also encouraged followers to voluntarily join a botnet to help promote and support their candidate, Matteo Salvini (Puente 2018). Activists and supporters could give their consent to the automatic publication of supporting messages on their Twitter wall by means of an application called 'LegaNordIllustrator'(Puente 2018). This delivery of messaging through authentic social media accounts is arguable and more insidious, as people are more likely to accept information coming from people they perceive to be familiar to them (Garrett & Weeks 2013). Through recruitment and engagement of people online, *Lega* could direct collective action aimed at pursuing its promotional campaign without violating the terms and conditions of social networks such as Twitter. As with The Movement, *Lega* also had supportive Facebook holdings that were found to be fake or in violation of the social network's community standards. One page, *Lega Salvini Premier Santa Teresa di riva*, shared a video which claims to be of migrants attacking a police car. The video, which is in fact from a movie, received some 10 million views (Avaaz 2019).

All three of these examples can be considered (mis)information operations as they are directed towards a target audience—the electoral constituency present online—with the aim of influencing public opinion through misleading activities (fake accounts) or content. Throughout, misleading and provocative content is used to fuel divides, particularly on issues related to migration, but also health. The network of accounts pushing this content is also vast and includes actors with different motivations, such as financial (Colono), religious (*La Luce di Mari*), and political (The Movement and *Lega*)—each intersecting where it is mutually beneficial, but at the same time recruiting and engaging authentic audiences to engage and spread content. A common element in the online holdings supporting these actors is a claim of being independent or grassroots. This approach might lead unsuspecting consumers to think that a given perspective is common to many different independent observers; and, in turn, this perspective might gain legitimacy. This risk is compounded when authentic accounts are coordinated to spread such material (Shin 2013).

## Discussion

What emerges from the example of Italy is that the line between truth and fiction is often blurred when dealing with misinformation. Misinformation is not only a matter of content, but also behaviour. For this reason, looking just at the content might not be sufficient for identifying and addressing (mis)information operations. Confirmation bias seems to lie at the core of the problem of misinformation--people tend to seize and frame their beliefs on specific information that reinforces their point of view (Schmidt, AL *et al.* 2018a; Del Vicario *et al.* 2016a). This, combined with a changing information environment in which anyone has quick access to any type of content almost costless, leads to the emergence of virtual echo chambers, groups of like-minded people clustering around a shared narrative, which shape, reinforce, and polarize users' beliefs. These polarized clusters of users seem particularly vulnerable to (mis)information operations, which—implementing a series of strategies and tactics aimed at promoting specific narratives—inevitably threaten the established social order and trigger feelings of distrust by leveraging what communities and social groups deeply fear.

As the matter is complex but absolutely crucial, the approach to addressing it needs to be cross-methodological and grounded on data-driven models by including psychometric and sociometric tests, massive data-analysis, as well as computational and network-based tools for both modelling and validation. To this respect, it may be useful to provide an integrated proposal for future research. Three main aspects seem to be worthy of future investigations: 1) the social-media environment, 2) the information dynamics, and 3) the use of automated systems for exploiting the cognitive and environmental vulnerabilities.

The social-media environment is obviously central. The dynamics of the social-media network structure can be characterized by steady rates of change, interrupted by sudden bursts. In particular, information diffusion in the form of cascades of post resharing often creates sudden bursts of new connections, which significantly reshape users' local network structure (Myers, Seth & Leskovec 2014). On the other side, the evolution of network structure, together with interdisciplinary approaches, might help to identify malicious/fake profiles (Conti, Hasani & Crispo 2013). Social media and social networks shape the debate on society and policy issues, but the dynamics of this process are not well understood. By monitoring social network activity on a range of issues, influential users and communities can be detected, communities can be classified according to their main interests, and sentiment analysis of the content can be performed to identify the leaning of each community towards a set of common topics and to identify controversial issues (Sluban *et al.* 2015; Kralj Novak, De Amicis & Mozetič 2018). To this respect, the main goal will be that of integrating the accumulated knowledge and lessons learned from the opinion polls with the new available social-media data sources (Grčar *et al.* 2017). Moreover, analyses that involve and compare different social media are needed in order to understand the universality of certain phenomena such as the presence of echo chambers, patterns of diffusion, and both information and misinformation.

Data science is emerging as a means to study information dynamics, including to quantitatively characterize and model the processes by which news pieces spread and are consumed, providing an early detection system for trends in public opinion. Take an anti-vaccine echo chamber, for example. It is now possible to identify which topics are attractive to users and the types of values they assign to them—in other words 'why not vaccinating is supposedly good'—and the beliefs underpinning their narrative. Based on that, it would be of interest to further investigate the possibility of detecting the informational cascades associated with a given topic and potentially neutralizing them. Unfortunately, carrying out an automated classification of misinformation remains difficult, particularly given the grey lines where most political opinion falls, but also due to the structural properties of content propagation (Conti *et al.* 2017). Another important aspect relates to understanding how the information environment affects the way people express their opinions and communicate, or to investigating whether the echo-chamber effect might also affect people's language other than people's behavior.

Regarding automated systems, social bot detection is an avenue of research. In the past years, bot detection techniques evolved from account-by-account techniques (Davis et al. 2016), where potential bots are compared individually against the elements of a single database of bots, to group analysis techniques, where groups of potential bots are jointly assessed through the patterns and regularities emerging from their coordinated behavior. The main reason behind such a shift is that social bots and botnets can be particularly sophisticated and almost indistinguishable from humans when analysed on a one-by-one basis (Cresci *et al.* 2017a). To this respect, two group-based detection techniques, namely the Social Fingerprinting (Cresci *et al.* 2017b) and Retweet-Buster (RTBust) (Mazza *et al.* 2019), have been proven to be particularly effective. In parallel to the shift towards group detection, bot detection techniques have also been evolving from supervised to unsupervised, and from reactive to proactive approaches (adversarial machine learning). In particular, a proactive approach starts with an existing model, and it simulates variations in the groups of accounts under examination. A new evaluation follows; and, if any of the automated accounts are left undetected, a threat is diagnosed. This, in turn, stimulates new design and sets out another research cycle. The next generation of techniques should strike a balance between accuracy, robustness, and generalizability, especially in light of the fact that the most insidious threats to the quality of online conversations might come from hybrid users (cyborgs), whose group coordination is less schematic as compared to completely automated accounts. Moreover—as bot and botnet detection lies at the convergence of three main communities, that is web and social media analysis, cyber intelligence and information security, analytics (data science, machine learning, artificial intelligence)—a comprehensive survey of the existing literature would be needed, while up-to-date and public datasets might allow more precise taxonomies of social bots and wide-ranging comparisons among detection techniques.

## Conclusion

This work addresses the problem of misinformation diffusion in the current information environment. This new disintermediated environment, dominated by social media, opens the door to (mis) information operations——that are the specific and organized use of misleading information to influence a target audience. The problem is particularly challenging for democracies, which derive their legitimacy on trust in information. The approach which has been adopted up to now seems to reduce the issue to a dichotomous distinction between falsity and veracity. We proposed to use an integrated and interdisciplinary approach to the problem, which takes into account the many facets of human cognitions and interactions as well as the complexity of the new information environment. In particular, we focused on the role played by confirmation bias, and on the relevance of specific personality traits, such as the narcissistic trait, in shaping the way in which social media users absorb and interact with information. Moreover, we described the ecosystem populating the new information environment and how it is manipulated through automated processes. We proceeded to explore (mis)information operations in the context of the Italian information environment, highlighting how a complex and holistic approach is necessary to correctly tackle the issue of emerging misinformation. Finally, an integrated proposal for future research has been provided focusing on three main aspects: the social-media environment, the information dynamics, and the use of automated systems for exploiting the cognitive and environmental vulnerabilities.

## Acknowledgement

This work was supported by the Slovenian Research Agency research core funding for the programme Knowledge Technologies (No P2-0103) and the European Union's Horizon 2020 research and innovation programme under grant agreement No 825153 - project EMBEDDIA (Cross-Lingual Embeddings for Less-Represented Languages in European News Media).

## References

Associated Press 2019, *EU election prompts removal of fake Italian Facebook accounts*, CBC, viewed 29 July 2019, <https://www.cbc.ca/news/technology/facebook-elections-1.5133720>.

Avaaz 2019, *Report: Far right networks of deception*, viewed 29 July 2019, <https://avaazimages.avaaz.org/Avaaz%20Report%20Network%20Deception%2020190522.pdf?slideshow>.

Bessi, A, Petroni, F, Del Vicario, M, Zollo, F, Anagnostopoulos, A, Scala, A, Caldarelli, G & Quat-trociocchi, W 2016, 'Homophily and polarization in the age of misinformation', *The European Physical Journal Special Topics*, vol. 225, no. 10, pp. 2047-59.

Bimber, B 2003, *Information and American democracy: Technology in the evolution of political power*, Cambridge University Press, Cambridge, MA, US.

Boczkowski, PJ, Mitchelstein, E, & Matassi, M 2018, '"News comes across when I'm in a moment of leisure": Understanding the practices of incidental news consumption on social media', *New Media & Society*, vol. 10, no. 20, pp. 3523-39.

Bond, RM, Fariss, CJ, Jones, JJ, Kramer, AD, Marlow, C, Settle, JE & Fowler, JH 2012, 'A 61-mil-lion-person experiment in social influence and political mobilization', *Nature*, vol. 489, no. 7415, p. 295.

Caplan, SE 2006, 'Relations among loneliness, social anxiety, and problematic Internet use', *CyberPsychology & Behavior*, vol. 10, no. 2, pp. 234-42.

Chak, K, & Leung, L 2004, 'Shyness and locus of control as predictors of Internet addiction and Internet use', *CyberPsychology & Behavior*, vol. 7, no. 5, pp. 559-70.

Chu, Z, Gianvecchio, S, Wang, H & Jajodia, S 2012, 'Detecting automation of Twitter accounts: Are you a human, bot, or cyborg?', *IEEE Transactions on Dependable and Secure Computing*, vol. 9, no. 6, pp. 811-24.

Cinelli, M, Brugnoli, E, Schmidt, AL, Zollo, F, Quattrociocchi, W & Scala, A 2019, 'Selective exposure shapes the Facebook news diet', *arXiv preprint arXiv:1903.00699*.

Conti, M, Hasani, A & Crispo, B 2013, 'Virtual private social networks and a Facebook implemen-tation', *ACM Transactions on the Web (TWEB)*, vol. 7, no. 3, p. 14.

Conti, M, Lain, D, Lazzeretti, R, Lovisotto, G & Quattrociocchi, W 2017, December, 'It's always April fools' day!: On the difficulty of social network misinformation classification via propagation features', *2017 IEEE Workshop on Information Forensics and Security (WIFS)*, IEEE, pp. 1-6.

Cresci, S, Di Pietro, R, Petrocchi, M, Spognardi, A & Tesconi, M 2017a, 'The paradigm-shift of social spambots: Evidence, theories, and tools for the arms race', *Proceedings of the 26th International Conference on World Wide Web companion*, International World Wide Web Conferences Steering Committee, pp. 963-72.

——2017b, 'Social fingerprinting: Detection of spambot groups through DNA-inspired behavioral modeling', *IEEE Transactions on Dependable and Secure Computing*, vol. 15, no. 4, pp. 561-76.

Davis, CA, Varol, O, Ferrara, E, Flammini, A & Menczer, F 2016, 'Botornot: A system to evaluate social bots', *Proceedings of the 25th International Conference Companion on World Wide Web*, International World Wide Web Conferences Steering Committee, pp. 273-4.

De Waal, E & Schoenbach, K 2010, 'News sites' position in the mediascape: Uses, evaluations and media displacement effects over time', *New Media & Society*, vol. 12, no. 3, pp. 477-96.

Del Vicario, M, Vivaldo, G, Bessi, A, Zollo, F, Scala, A, Caldarelli, G, & Quattrociocchi, W 2016a, 'Echo chambers: Emotional contagion and group polarization on Facebook', *Scientific Reports*, vol. 6, no. 37825.

Del Vicario, M, Bessi, A, Zollo, F, Petroni, F, Scala, A, Caldarelli, G, Stanley, HE & Quattrociocchi, W 2016b, 'The spreading of misinformation online', *Proceedings of the National Academy of Sciences*, vol. 113, no. 3, pp. 554-59.

Del Vicario, M, Gaito, S, Quattrociocchi, W, Zignani, M, & Zollo, F 2017, 'Public discourse and news consumption on online social media: A quantitative, cross-platform analysis of the Italian Referendum', *arXiv preprint arXiv:1702.06016*.

Del Vicario, M, Quattrociocchi, W, Scala, A, & Zollo, F 2019, 'Polarization and fake news: Early warning of potential misinformation targets', *ACM Transactions on the Web (TWEB)*, vol. 13, no. 2, p. 10.

Ebner, J & Davey, J 2018, *Mainstreaming Mussolini: How the extreme right attempted to 'Make Italy Great Again' in the 2018 Italian election*, Institute for Strategic Dialogue (ISD), viewed 29 July 2019, <https://www.isdglobal.org/wp-content/uploads/2018/03/Mainstreaming-Mussoli-ni-Report-28.03.18.pdf>.

Eksi, F 2012, 'Examination of narcissistic personality traits: Predicting level of Internet addiction and cyber bullying through path analysis', *Educational Sciences: Theory and Practice*, vol. 12, no. 3, pp. 1694-706.

Epstein, R & Robertson RE 2015, 'The Search Engine Manipulation Effect (SEME) and its possible impact on the outcomes of elections', *Proceedings of the National Academy of Sciences*, vol. 33, no. 112, pp. E4512-21.

Garrett, RK & Weeks, BE 2013, 'The promise and peril of real-time corrections to political misperceptions', Proceedings of the 2013 conference on Computer Supported Cooperative Work, ACM, pp. 1047-58.

Gavazzi L 2019 (IT), *Elezioni politiche 2018: Sondaggi e previsioni seggi,* (Political elections 2018: polls and forecasts), *Panorama*, viewed 29 July 2019, <https://www.panorama.it/news/po-litica/elezioni-politiche-italiane-2018-sondaggi/>.

Grčar, M, Cherepnalkoski, D, Mozetič, I & Kralj Novak, P 2017, 'Stance and influence of Twitter users regarding the Brexit referendum', *Computational Social Networks*, vol. 4, no. 1, p. 6.

Grinberg, N, Joseph, K, Friedland, L, Swire-Thompson, B, & Lazer, D 2019, 'Fake news on Twitter during the 2016 US presidential election', *Science*, vol. 363, no. 6425, pp. 374-8.

Guess, A, Nagler, J & Tucker, J 2019, 'Less than you think: Prevalence and predictors of fake news dissemination on Facebook', *Science Advances*, vol. 5, no. 1, eaau4586.

Haselton, MG, Nettle, D & Murray, DR 2015, 'The evolution of cognitive bias', *The handbook of evolutionary psychology*, Wiley, n.p., pp. 1-20.

Jiang, M, Cui, P & Faloutsos, C 2016, 'Suspicious behavior detection: Current trends and future directions', *IEEE Intelligent Systems*, vol. 31, no. 1, pp. 31-9.

Joint Chiefs of Staff (JCS) 2014, *Joint doctrine for information operations, Joint Pub 3-13*, viewed 29 July 2019, <https://fas.org/irp/doddir/dod/jp3_13.pdf>.

Kim, EJ, Namkoong, K, Ku, T & Kim, SJ 2008, 'The relationship between online game addiction and aggression, self-control and narcissistic personality traits', *European Psychiatry*, vol. 23, no. 3, pp. 212-8.

Kralj Novak, P, De Amicis, L & Mozetič, I 2018, 'Impact investing market on Twitter: Influential users and communities', *Applied Network Science*, vol. 3, no. 1, p. 40.

Macdonald, S 2006, *Propaganda and information warfare in the Twenty-First Century: Altered images and deception operations*, Routledge, London, UK.

Mackay, J 2019, 'How Italy's 'digital populists' used the anti-vaccine agenda to propel themselves into power', *Political Critique*, viewed 29 July 2019, <http://politicalcritique.org/world/eu/2019/how-italys-digital-populists-used-the-anti-vaccine-agenda-to-propel-themselves-into-power/>.

Marino, C, Finos, L, Vieno, A, Lenzi, M & Spada, MM 2017, 'Objective Facebook behaviour: Differences between problematic and non-problematic users', *Computers in Human Behavior*, vol. 73, pp. 541-6.

——, Mazzieri, E, Caselli, G, Vieno, A & Spada, MM 2018, 'Motives to use Facebook and problematic Facebook use in adolescents', Journal of Behavioral Addictions, vol. 7, no. 2, pp. 276-83.

——, Marci, T, Ferrante, L, Alto, G, Vieno, A, Simonelli, A, Caselli, G & Spada, MM 2019a, 'Attachment and metacognitions as predictors of problematic Facebook use in adolescents', *Journal of Behavioral Addictions*, vol. 8, no. 1, pp. 63–78.

——, Caselli, G, Lenzi, M, Monaci, MG, Vieno, A, Nikčević, AV & Spada, MM 2019b, 'Emotion regulation and desire thinking as predictors of problematic Facebook use', *Psychiatric Quarterly*, vol. 90, no. 2, pp. 405-11.

Mazza, M, Cresci, S, Avvenuti, M, Quattrociocchi, W & Tesconi, M 2019, 'RTbust: Exploiting temporal patterns for botnet detection on Twitter', *arXiv preprint arXiv:1902.04506*.

Mehdizadeh, S 2010, 'Self-presentation 2.0: Narcissism and self-esteem on Facebook'. *Cyberpsychology, Behavior, and Social Networking*, vol. 13, no. 4, pp. 357-64.

Mocanu, D, Rossi, L, Zhang, Q, Karsai, M & Quattrociocchi, W 2015, 'Collective attention in the age of (mis) information', *Computers in Human Behavior*, vol. 51, pp. 1198-204.

Mustafaraj, E & Metaxas, PT 2010, 'From obscurity to prominence in minutes: Political speech and real-time search', *WebSci10: Extending the frontiers of society on-line*.

Myers, SA, & Leskovec, J 2014, 'The bursty dynamics of the Twitter information network', *Proceedings of the 23rd International Conference on World Wide Web*, ACM, pp. 913-24.

Nardelli A, Silverman C 2016, 'Italy's most popular political party is leading Europe in fake news and Kremlin propaganda', *BuzzFeed*, viewed 29 July 2019, <https://www.buzzfeed.com/albertonardelli/italys-most-popular-political-party-is-leading-europe-in-fak>.

——2017, 'One of the biggest alternative media networks in Italy is spreading anti-immigrant news and misinformation on Facebook', *BuzzFeed*, viewed 29 July 2019, <https://www.buzzfeed. com/albertonardelli/one-of-the-biggest-alternative-media-networks-in-italy-is>.

Nickerson, RS 1998, 'Confirmation bias: A ubiquitous phenomenon in many guises', *Review of General Psychology*, vol. 2, no. 2, pp. 175-220.

Pomerantsev, P, Appelbaum A, Gaston, S, Fusi, N, Peterson Z, Quattrociocchi, W, Zollo F, Schmidt, AL, Peruzzi, A, Severgnini, B, de Cesco AF & Casati, D 2019, *Journalism in the age of populism and polarisation: Insights from the migration debate in Italy*, LSE Arena, viewed 29 July 2019, <http://www.lse.ac.uk/iga/assets/documents/arena/archives/Journalism-In-The-Age-of-Populism-and-Polarisation.pdf>.

Puente D 2018, 'I tweet *automatici della Lega Nord su* Twitter: "*Un Matteo #Salvini Strepitoso su LA7! Siete d'accordo?*"', ('*Lega-nord* Automatic tweets on Twitter: "A Sensational Matteo

#Salvi-ni"'), *Il Blog di* David Puente, viewed 29 July 2019, <https://www.davidpuente.it/blog/2018/01/24/tweet-automatici-della-lega-nord-su-twitter-un-matteo-salvini-strepitoso-su-la7-siete-daccordo/>.

Quattrociocchi, W 2017, 'Inside the echo chamber', *Scientific American*, vol. 316, no. 4, pp. 60-3.

Ruths, D 2019, 'The misinformation machine', *Science*, vol. 363, no. 6425, pp. 348.

Schmidt, AL, Zollo, F, Del Vicario, M, Bessi, A, Scala, A, Caldarelli, G, Stanley, HL & Quattrociocchi, W 2017, 'Anatomy of news consumption on Facebook', *Proceedings of the National Academy of Sciences*, vol. 114, no. 12, pp. 3035-9.

Schmidt, AL, Zollo, F, Scala, A & Quattrociocchi, W 2018a, 'Polarization rank: A study on European news consumption on Facebook', *arXiv preprint arXiv:1805.08030.*

Schmidt, AL, Zollo, F, Scala, A, Betsch, C, & Quattrociocchi, W 2018b, 'Polarization of the vaccination debate on Facebook', *Vaccine*, vol. 36, no. 25, pp. 3606-12.

Schmidt, C 2019, 'Facebook is committing $300 million to support news, with an emphasis on local', *NiemanLab*, viewed 29 July 2019, <https://www.niemanlab.org/2019/01/facebook-is-committing-300-million-to-support-news-with-an-emphasis-on-local/>.

Shin, DH 2013, 'User experience in social commerce: In friends we trust', *Behaviour & Information Technology*, vol. 32, no.1, pp. 52-67.

Sluban, B, Smailović, J, Battiston, S & Mozetič, I 2015, 'Sentiment leaning of influential communities in social networks', *Computational Social Networks*, vol. 2, no. 1, p. 9.

Spada, MM & Marino, C 2017, 'Metacognitions and emotion regulation as predictors of problematic Internet use in adolescents', *Clinical Neuropsychiatry*, vol. 14, no. 1, pp. 59-63.

Stieglitz, S, Brockmann, T & Dang-Xuan, L 2012, 'Usage of social media for political communication', *PACIS*, July, p. 22.

Sunstein, CR 2002, 'The law of group polarization', *Journal of Political Philosophy*, vol. 10, no. 2, pp. 175-95.

——2018, *# Republic: Divided democracy in the age of social media*, Princeton University Press.

VandenBos, GR 2007, *APA dictionary of psychology,* American Psychological Association, viewed 29 July 2019, <https://dictionary.apa.org/cognitive-psychology>.

Vosoughi, S, Roy, D & Aral, S 2018, 'The spread of true and false news online', *Science*, vol. 359, no. 6380, pp. 1146-51.

Wanless, A & Berk, M 2019, 'The audience is the amplifier: Participatory propaganda', *The SAGE handbook of propaganda*, Sage, London, UK.

World Economic Forum (WEF) 2017, *The global risks report 2017*, viewed 29 July 2019, <http://www3.weforum.org/docs/GRR17_Report_web.pdf>.

Zanotto, L 2017, '*Caratteristiche psicologiche delle reti sociali su Facebook*' ('Psychological characteristics of social networks on Facebook'), Master's thesis, *Università degli Studi di Padova* (University of Padua), Padua, Veneto, IT.

Zollo, F, Bessi, A, Del Vicario, M, Scala, A, Caldarelli, G, Shekhtman, L, Havlin, S & Quattrociocchi, W 2017, 'Debunking in a world of tribes', *PloS One*, vol. 12, no. 7, e0181821.

Zollo, F, Novak, PK, Del Vicario, M, Bessi, A, Mozetič, I, Scala, A & Quattrociocchi, W 2015, 'Emotional dynamics in the age of misinformation', *PloS One*, vol.12, no. 9.